\documentclass[doublecol]{epl2} 
\usepackage{xcolor}
\colorlet{rev}{black}

\title{Cultural heterogeneity constrains diffusion of innovations}
\shorttitle{Cultural heterogeneity constrains diffusion of innovations} %Insert here a short version of the title if it exceeds 70 characters

\author{Aruane M. Pineda\inst{1} \and Sandro M. Reia\inst{2} \and Colm Connaughton \inst{3,4} \and José F. Fontanari \inst{5} \and Francisco A. Rodrigues \inst{1}}
\shortauthor{Pineda AM \etal}

\institute{                    
\inst{1} Institute of Mathematical and Computer Sciences, University of São Paulo, São Carlos, São Paulo, Brazil\\
\inst{2} Lyles School of Civil Engineering, Purdue University, West Lafayette, Indiana, USA \\
\inst{3} London Mathematical Laboratory, London, England, UK \\
\inst{4} Mathematics Institute, University of Warwick, Coventry, England, UK \\
\inst{5} São Carlos Institute of Physics, University of São Paulo, São Carlos, São Paulo, Brazil
}

\abstract{
Rogers' diffusion of innovations theory asserts that the cultural similarity among individuals plays a crucial role on the acceptance of an innovation in a community. 
However, most studies on the diffusion of innovations have relied on epidemic-like models where the individuals have no preference on  whom they interact with. Here, we use an agent based model to study the diffusion of innovations in a community of synthetic heterogeneous agents whose interaction preferences depend on their cultural similarity. The community heterogeneity and the agents' interaction preferences are described by Axelrod's model, whereas the diffusion of innovations is described by a variant of the Daley and Kendall  model of rumour propagation. 
The interplay between the social dynamics and the spreading of the innovation  is controlled by the parameter $p \in [0,1]$, which yields the probability that the agent  engages in social interaction or attempts to spread  the innovation. 
Our findings support Roger's empirical observations that cultural heterogeneity curbs the diffusion of innovations.}

\begin{document}

\maketitle

\section{Introduction}

Everett Rogers proposed in 1962 a theory for the diffusion of innovations that explained  how novel ideas spread and are perceived by individuals in a community \cite{thirdedition}. Among the goals of Rogers' ambitious enterprise  was the  identification of the elements that determine how innovations spread among individuals, communities, societies, and nations \cite{rogers2014diffusion,wejnert2002integrating}. 
The social fabric, or the relationships and connections that individuals make with each other, is one of these elements:  Rogers found that the degree of cultural heterogeneity of a community  influences how far information spreads among its members, concluding that cultural similarities facilitate the diffusion of innovations.  

Individual heterogeneity refers to the variety of traits and behaviors that characterize the individuals within a community. Individuals that exhibit a wide range of traits and behaviors may also have varying requirements and preferences that can affect both their interests  and the rate of adoption of novelties. Individual heterogeneity can result in a more varied and effective information flow, which can speed up the diffusion of innovations. The degree of shared values, beliefs, and norms among members of a community is referred to as cultural similarity \cite{thirdedition}. People of similar cultural backgrounds may have similar requirements and preferences.
%, which helps increase comprehension of and acceptance of new developments. 
According to Rogers, cultural similarity can speed up the spreading of innovations by lowering uncertainty and fostering interpersonal trust \cite{dosi1988nature,thirdedition, petrakis2016cultural}.

To facilitate mathematical analysis, individual heterogeneity is usually not taken into account in the models for the propagation of rumours, innovations, ideas, or even infectious diseases~\cite{Arruda2018, ferraz22}. However,
individual heterogeneity and cultural similarity can  easily be incorporated in those models by borrowing the  agents' representation and the  interaction rules introduced by Axelrod in his celebrated model of culture dissemination \cite{axelrod1997dissemination}. 
Axerold's model offers an explanation for the fact that although social interactions increase the similarity between individuals -- a phenomenon known as social influence -- there is still cultural diversity in human societies. In a two-dimensional lattice where the agents interact with their nearest neighbors only,  the model exhibits two classes of absorbing configurations: ordered (monocultural) configurations in which all  agents (or the vast majority of them) exhibit the same culture, and disordered (multicultural) configurations in which multiple cultures coexist in the lattice. 
The absorbing configuration is the result of the interplay between the disorder of the initial configuration, tuned by the two parameters of the model, viz., $F$ (number of  cultural features) and $Q$ (number of values each cultural feature takes on), and social influence \cite{castellano2000nonequilibrium,klemm2003nonequilibrium}. Interestingly, the model shows that multicultural configurations  are possible outcomes of a dynamics where agents become more alike after each interaction.

The diffusion of  new ideas in the academic world has been modeled using  SIR-like epidemic models \cite{bettencourt2006power,reia2021sir}, but the standard epidemic model for studying the spreading of  rumours is the DK model proposed by Daley and Kendall in the 1960s \cite{daley1964epidemics}.
In the DK model, the agents can assume three possible states, viz., $S$ (ignorant), $I$ (spreader) and $R$ (stifler).
Ignorants  are not aware of the rumour, spreaders are aware of the rumour and spread it, whereas stiflers are aware of the rumour but do not  spread it.

\begin{figure}
\includegraphics[width=0.45\textwidth]{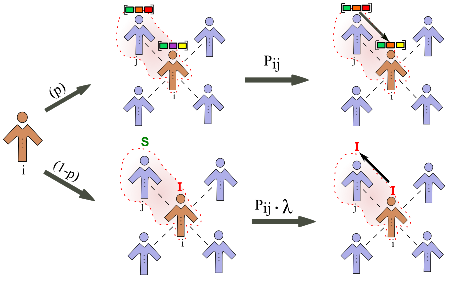}
\caption{\textbf{Mechanisms of cultural assimilation and rumour propagation.}
Agent $i$ takes part in the cultural dynamics with probability $p$ and in the spreading of the rumour with probability $1 - p$.
In the cultural dynamics, agent $i$ selects a random neighbor $j$ and assimilates a divergent feature with probability $P_{ij}$.
In the rumour dynamics, partially represented here, a spreader agent $i$ transmits the rumour to an ignorant neighbor $j$ with probability $\lambda P_{ij}$.}
\label{fig_illustration}
\end{figure}

\begin{figure*}[t!]
\includegraphics[width=0.975\textwidth]{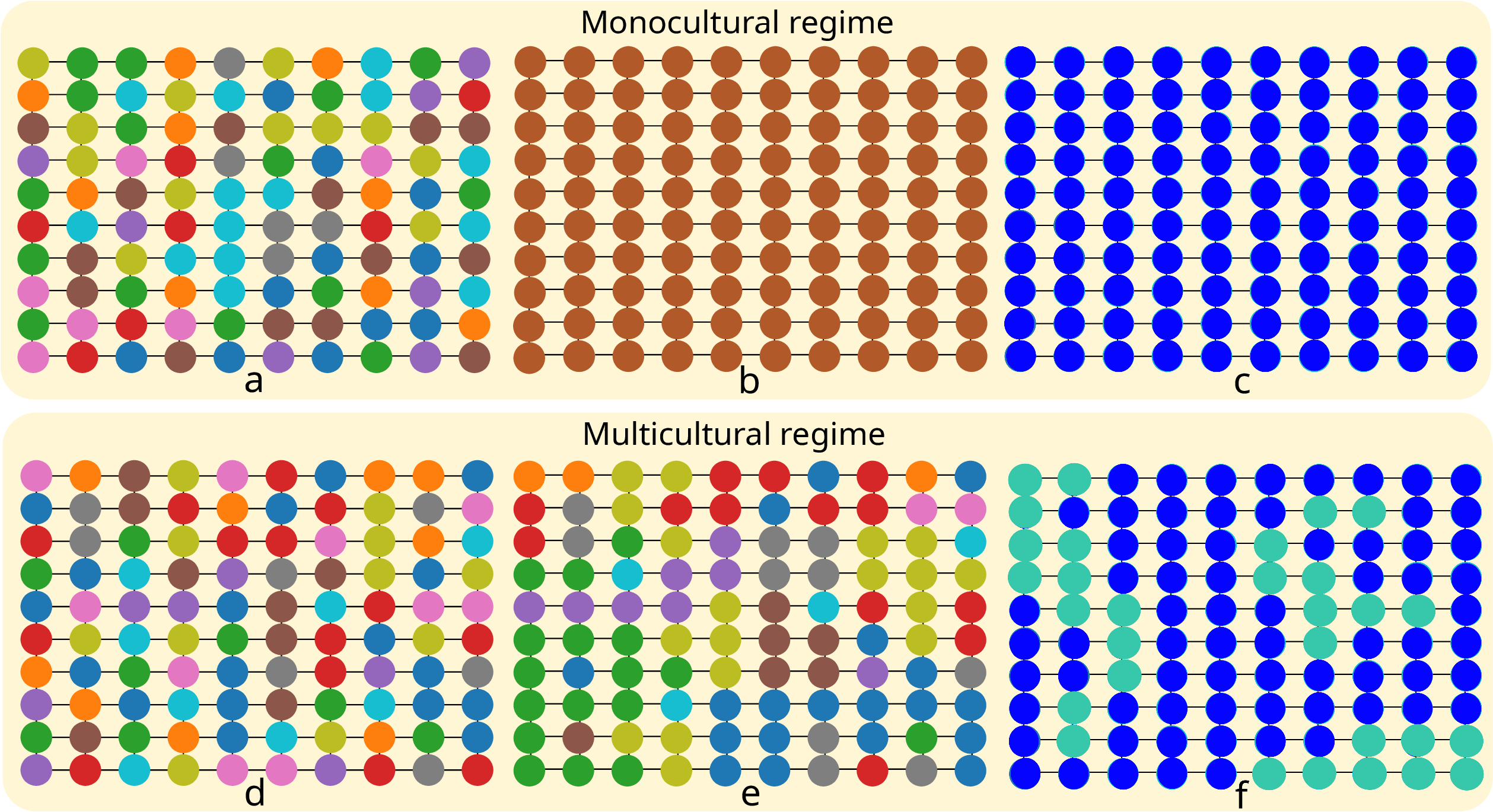}
\caption{\textbf{Absorbing configurations of the coupled culture-rumour  model.}
\textcolor{rev}{
%Cultural heterogeneity drives the dynamics toward a monocultural or multicultural steady state.
For a fixed number of cultural features $F$, the  diversity of the community is captured by the number of cultural traits $Q$.
In a scenario of low cultural diversity, interactions increase even further the similarity between agents, which in turn facilitates the spreading of the rumour.
For  highly diverse  communities, the steady state is characterized by the presence of many distinct cultural domains, which  hamper the spreading of the rumour.
}
The top row shows the initial (a) and final (b) cultural states of the agents for $F = 3$ and $Q = 4$.
Each color corresponds to a cultural vector, so we can observe the ordering effect of social influence on the disordered initial configuration.
Panel (c) shows that all agents became stiflers (blue) at the final state when there is a single spreader in the initial configuration.
The bottom row shows the initial (d) and final (e) cultural states of the agents for $F = 3$ and $Q = 14$. In this case, social influence is not strong enough to homogenize the cultural states of the agents. Panel (f) shows that the rumour spreads across different cultural domains, but some ignorants (light blue) remain in the community. 
The other parameters are  $L = 10$, $\lambda = 1$, $\alpha = 0.01$ and $p = 0.1$.}
\label{fig_mono_multi_states}
\end{figure*}

%In this study, we aim to examine how innovations spread in a system with heterogeneous agents providing more evidence to support the Everett Rogers hypothesis.
\textcolor{rev}{
Here, we  combine the Axelrod model and the DK model to study the diffusion of innovations (i.e., rumours) in a more realistic scenario where the agents are culturally heterogeneous at the outset, but their cultural dissimilarities change dynamically as they interact. This approach allows us to further investigate Rogers' empirical observations regarding the relationship between cultural heterogeneity and the success of innovations.}
%By using a complex's system methodology, we combine the Axelrod model and the DK model to study the spreading of an innovation (i.e., a rumour) in a community of heterogeneous agents. 
%In this study, our goal is to explore the complex relationship between diffusion of innovations (i.e., a rumour) and individual heterogeneity by combining the Axelrod model with the DK model.
%This coupling allows us to model the diffusion of innovations in a more realistic scenario where individuals are heterogeneous and cultural similarities between them change dynamically as they interact.
%Here we combine the Axelrod model and the DK model to study the spreading of an innovation (i.e., a rumour) in a community of heterogeneous agents. 
The coupling between these models is that the spreader's attempt to pass on the rumour to an ignorant or to a stifler is conditioned on the cultural similarity between those agents.  
In that sense, our approach differs starkly from previous studies of diffusion of innovations in Axelrod's model in which the innovation was a novel value of a cultural feature \cite{Tilles2015, reia2020diffusion}. The cultural or Axelrod's dynamics takes place with probability $p$, whereas  the rumour or DK  dynamics takes place with probability $1 - p$, so $p$ can be viewed as a proxy for the openness of agents to social influence. A similar idea was used to verify the impact of rumour propagation on disease spreading~\cite{Ventura19, Velasquez2020} and to study interacting diseases~\cite{Ventura2021}.  Henceforth we will use the words innovation and rumour interchangeably. 

Since we keep the number of cultural features fixed to $F=3$ throughout  this paper, the cultural diversity of the community is controlled solely by the parameter $Q$: the greater the number of distinct values that each cultural feature can take on, the more  cultural diverse the community is. %Henceforth we will refer to $Q$ as the cultural diversity parameter. 
The main focus of this study is the fraction of agents that are aware of the rumour after the dynamics reaches an absorbing configuration\textcolor{rev}{, which is given by the ratio between the number of stiflers and the community size. We note that in an absorbing configuration there can be only stiflers and ignorants.  Henceforth, we will  denote the average of this ratio over many independent runs by $\langle R \rangle$. 
Clearly, $\langle R \rangle$ is a proxy for the success of the rumour: the greater it is, the more agents are aware of the rumour. The cultural diversity $Q$ and the openness to social influence $p$ will be the leading independent variables in our study.
In particular, we find that $\langle R \rangle$ decreases with increasing $Q$ and increases with increasing $p$.} Therefore,  heterogeneity makes it difficult for a rumour to spread in a community and so different narratives may emerge without need of topological changes \cite{holme2006nonequilibrium, fu2008coevolutionary, iniguez2009opinion}.

\section{Methods}
The agent based model we study here is illustrated in Fig.~\ref{fig_illustration}.
%The heterogeneity dynamics given by Axelrod's model determines the cultural states of the agents, which in turn modulates the propagation of rumours given by a version of   Maki and Thompson ($MT$) model \cite{maki1973mathematical}.
Agents are placed in a square lattice of side $L$ with periodic boundary conditions and neighborhood given by the four nearest sites. As mentioned before, the Axelrod and the rumour dynamics take place with probability $p$ and $1 - p$, respectively. The cultural and the rumour dynamics are, in principle, independent. In the following, we present the details of the implementation of both dynamics.

\subsection{The Axelrod Model}

Each agent $i$ is assigned a cultural vector $\Phi_i =(\phi_{i1}, \phi_{i2}, . . ., \phi_{iF })$ of length $F$.
Each component of this vector represents a cultural feature prone to social influence, and each feature can take on $Q$ values (traits), i.e.,  $\phi_{ik} =\{1, \ldots, Q\}$ \cite{axelrod1997dissemination}.
The number of features ($F$) and traits ($Q$) determine the degree of disorder of the initial configuration since there are $Q^F$ possible cultural states (or cultures) for each agent. The cultural dynamics proceeds as follows:

\begin{itemize}
\item An agent $i$ is selected at random.
\item An agent $j$ $\in$ $V_{i}$, where $V_{i}$ is the set of neighbors of agent $i$, is selected at random.
\item With  probability given by the cultural similarity of agents $i$ and $j$, viz.,
\begin{equation}\label{pij}
   P_{ij} = \frac{1}{F}\sum_{k = 1}^F \delta_{\phi_{i k}, \phi_{j k}}, 
\end{equation}
a feature $k$ such that  $\phi_{ik} \neq \phi_{jk}$ is selected at random and $\phi_{jk} \rightarrow \phi_{ik}$. Here $\delta_{\phi_{i k}, \phi_{j k}} = 1$ if $\phi_{ik} =  \phi_{jk}$ and $0$ otherwise
\end{itemize}

\noindent 
%The random choices are obtained from a uniform distribution.
These steps are repeated until an absorbing configuration is reached. In an absorbing configuration, any pair of neighbouring  agents must share no cultural features (i.e., $P_{ij} = 0$) or have identical cultures (i.e., $P_{ij} = 1$). 

\begin{figure*}[h]
\includegraphics[width=1\textwidth]{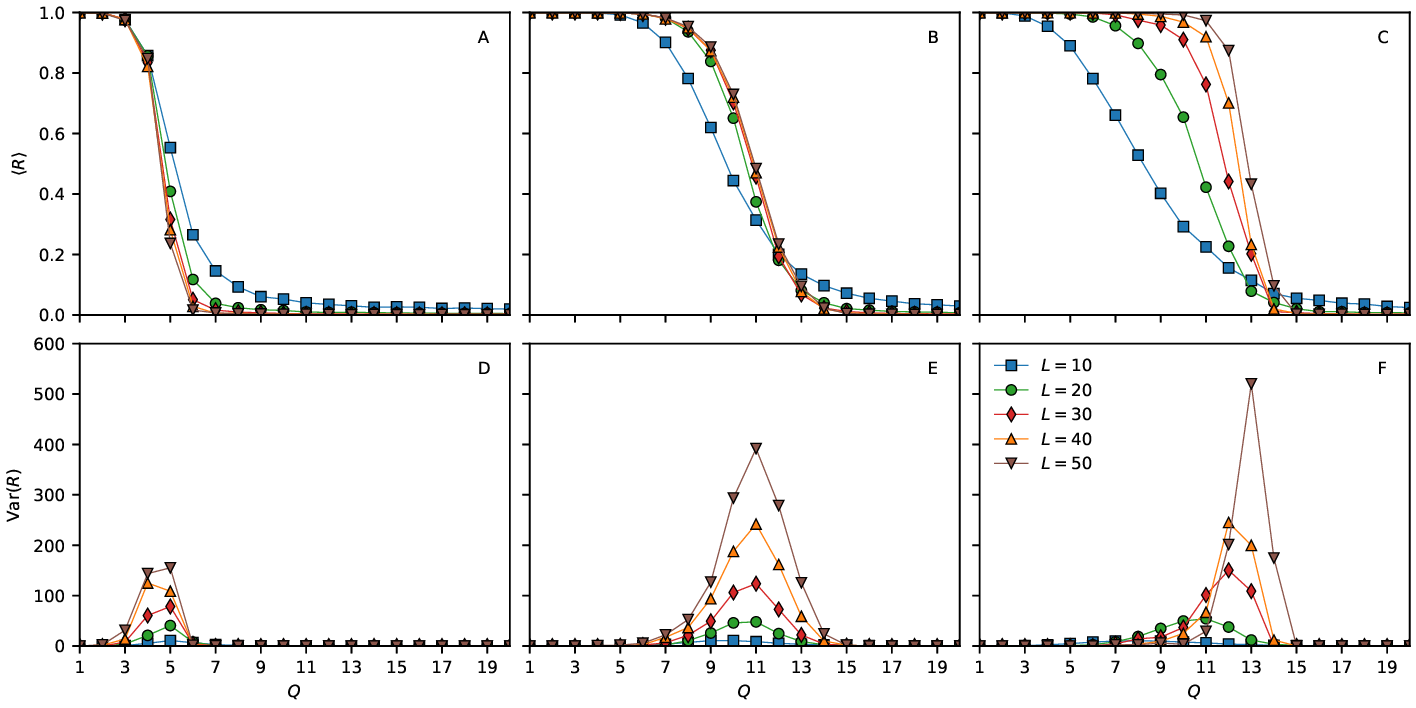}
\caption{
\textbf{Competition of cultural diversity and openness to social influence  on the spreading of rumours.}
\textcolor{rev}{
%As the openness to social influence increases from $p = 0$ (panels A and D) to $p = 0.5$ (panels B and E) and $p = 1.0$ (panels C and D), the transition peak of the mean number of stiflers ($\langle R \rangle$) shifts to the right (to higher values of $Q$).
Increase of cultural diversity $Q$ decreases the mean fraction of stiflers $\langle R \rangle$, whereas increase of  openness to social influence $p$ decreases $\langle R \rangle$.  
%the range of $Q$ in which the mean fraction of stiflers $\langle R \rangle \approx 1$ increases, indicating that the rumour spreads more easily when agents are more open to cultural exchange.
}
%Mas function of the cultural diversity $Q$. 
Panel A: $p = 0$, i.e., the culture of the agents is frozen in the initial random configuration.
Panel  B: $p = 0.5$. Panel C: $p = 1$, i.e., the rumour propagates only after the cultural dynamics reaches an absorbing configuration.
%Each curve was obtained in a system with different size $L$ with $F = 3$, $\lambda = 1.00$, and $\alpha = 0.01$. 
Panels D, E and F  show the variances of the  fraction of stiflers  for $p=0, 0.5$ and $1$, respectively.  The peak of the variance for large $L$ yields the approximate location of the transition point between the regimes $\langle R \rangle \approx 1$ and $\langle R \rangle \approx 0$. The lattice sizes $L$ are  indicated in Panel F. The other parameters are  $F = 3$, $\lambda = 1$ and $\alpha = 0.01$.}
\label{fig_results}
\end{figure*}

\subsection{The Rumour Model}

Here we use the variant of the DK model introduced by Maki and Thompson~\cite{maki1973mathematical}, which we will refer to as MT model.
Whereas in the DK model both agents can have their states updated in a single interaction, in the MT model only one agent is  updated per interaction.
Despite this difference, the outcomes of both models  are similar and well described by the same mean-field approach, with the DK model exhibiting a higher dispersion of mean values than the MT model \cite{doi:10.1137/100819588}. 
In addition, in both rumour models the agents are homogeneous, in the sense that they have ({\em i}) the same probability of receiving and re-transmitting the rumour when they become  spreaders, and ({\em ii}) the same probability of becoming  stiflers.

The spreading of the rumour is a contact process driven by the following  rules:

\begin{itemize}
\item An agent $i$ is selected at random.
\item If agent $i$ is a spreader, then: 
\begin{itemize}
\item An agent $j \in V_i$ is selected at random.
\item If agent $j$ is ignorant, then agent $j$ becomes a spreader with probability $\lambda P_{i,j} $.
\item If agent $j$ is a spreader or a stifler, then agent $i$ becomes a stifler with probability $\alpha P_{i,j}$.
\end{itemize}
\end{itemize}

\noindent These steps are repeated until the dynamics reaches an absorbing configuration where there are no spreaders in the lattice.
To ensure convergence, we  introduce the recovery probability $\delta= 10^{-6}$ for a spreader to turn into a stifler spontaneously \cite{nekovee2007theory}.

We note that the propagation of the rumour is modulated by the cultural similarity  of the interacting agents. In other words, the propagation of the rumour from agent $i$ to agent $j$ is possible if and only if these agents have at least one cultural feature in common to trigger a social interaction (i.e., $P_{ij} > 0$). The cultural dynamics, however,  is not influenced by the rumour dynamics.

\section{Results}

The  mean fraction of stiflers $\langle R \rangle$ in the absorbing configurations is frequently used as an order parameter in epidemic models \cite{kermack1927contribution} and so here we use this measure to characterize the asymptotic regime of the coupled culture-rumour dynamics. As already mentioned,  the cultural dynamics is controlled by the parameters $F$ and $Q$, whereas the (pure) rumour dynamics is controlled by  $\lambda$ and $\alpha$.  For fixed $F=3$,  increase of $Q$ leads to a sharp transition from a monocultural to a multicultural regime \cite{castellano2000nonequilibrium,klemm2003nonequilibrium}, as illustrated in  Fig. \ref{fig_mono_multi_states}.

\begin{figure}
\includegraphics[width=0.45\textwidth]{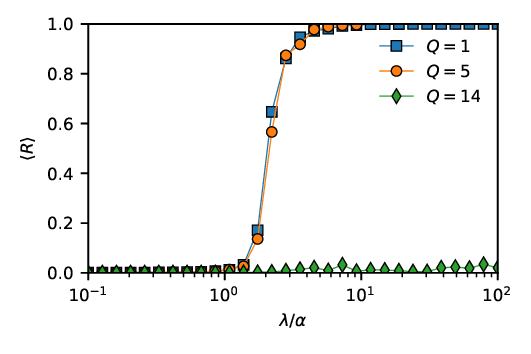}
\caption{\textbf{Effect of the basic reproduction number on the spreading of rumours.}
Mean fraction of stiflers $\langle R \rangle$ as function of the basic reproduction number $\mathcal{R}_0 =\lambda / \alpha$ for  $Q=1, 5$ and $14$ as indicated.
\textcolor{rev}{
In low diversity communities ($Q = 1$ and $Q = 5$), the rumour spreads over a sizeable portion of the community provided that  $\mathcal{R}_0> 1$, but for high diversity communities ($Q = 14$) it is limited to a tiny region around its creator, regardless of the value of $\mathcal{R}_0$. 
}
The other parameters are  $F = 3$, $L = 50$, $p = 0.5$ and $\alpha = 0.01$.
We keep $\alpha$ fixed and vary $\lambda$ to change the ratio $\lambda / \alpha$. 
}
\label{Fig_R0}
\end{figure}

Figure \ref{fig_results} shows that $\langle R \rangle$  is strongly influenced by the cultural heterogeneity of the community. We  observe two asymptotic regimes: a regime where the rumour  spreads to a sizeable portion of the lattice (i.e., $ \langle R  \rangle \approx 1$), and a regime where the rumour is restricted to a microscopic portion of the lattice (i.e., $ \langle R  \rangle \approx 0$). The transition point between  these regimes depends on the cultural and epidemiological parameters of the coupled culture-rumour  dynamics. However, since our primary interest is  the effect of cultural heterogeneity on the propagation of the rumour, we set $\lambda = 1$ and $\alpha = 0.01$, which correspond to the  basic reproduction number $\mathcal{R}_0 = \lambda / \alpha = 100 $. This guarantees  that the rumour spreads to the entire lattice ($\langle R \rangle = 1$) in  the case of homogeneous agents (i.e, for $Q=1$) as can be seen in the top panels of Fig. \ref{fig_results}.

These results  show that $\langle R \rangle$ decreases as $Q$ increases,  regardless of value of the parameter $p$, which measures the openness of the agents to social influence. Thus, cultural heterogeneity always impairs the spreading of the rumour. As $p$ increases from $p = 0$ to $p = 1$, the curves 
$\langle R \rangle$ {\em vs.} $Q$ shift to the right (top panels of 
Fig. \ref{fig_results}), revealing that the spreading of the rumour is  facilitated by the openness to social influence.
%(Fig. \ref{fig_results}A) to $p=0.5$ (Fig. \ref{fig_results}B) and then $p = 1.0$ (Fig. \ref{fig_results}C), the transition curve shifts to the right, revealing that rumour spreads more easily when agents are more open to social influence. Interestingly, the transition curve shifts to left when $p \approx 1$. 

In fact,  recalling that for $p = 0$ the rumour dynamics takes place in the frozen initial configuration of the agents' cultures, which corresponds to the most heterogeneous (or disordered) configuration for a given $F$ and $Q$. Consequently, we observe a very small range of $Q$ where  $\langle R \rangle \approx 1$, indicating that the spreading of the rumour is  heavily impaired by cultural heterogeneity.
On the other hand, for $p = 1$  the cultural dynamics proceeds until it reaches an absorbing configuration, and only then the rumour dynamics sets in. In this case, 
we observe a wider range of $Q$ where $\langle R \rangle \approx 1$, which highlights  that  openness to social influence facilitates the spread of rumours.

%The transition point $Q_c(L)$ might also depend on $p$. Figure \ref{fig_results} shows the effects of the limit cases in which $p = 0$ and $p = 1$ on the phase transition of the system.  

A word is in order about the phase transition between the asymptotic regimes characterized by $\langle R \rangle > 0$ and $\langle R \rangle \to 0$.  Since a phase transition occurs only in the thermodynamic limit $L \to \infty$, we can only infer its onset for finite systems by considering lattices of different sizes \cite{landau1990monte}, as done in  Fig. \ref{fig_results}. A typical sign of the onset of a phase transition is the increase of the size of the fluctuations, which in our case is captured by the variance of the fraction of stiflers. Thus, the value $Q = Q_c(L)$ at which the variance is maximum offers a rough estimate of 
the  transition point, as illustrated  in the bottom panels of  Fig. \ref{fig_results}.

It is interesting that for $p=0$ and $\mathcal{R}_0 \to \infty$ our problem reduces to the bond percolation in the square lattice \cite{percolation} where the probability that a bond transmits the fluid is $1 - \left ( 1 -1/Q \right )^F$ and so $Q_c = \left ( 1 - 2^{-1/F} \right )^{-1}$. For $F=3$, it yields $Q_c \approx 4.85$ that agrees very well with the results of panel A of Fig. \ref{fig_results}. More importantly, since in the bond percolation we have $0 < \langle R \rangle < 1$ at the critical point $Q=Q_c$, the transition is discontinuous and only for $Q=1$ (homogeneous agents) the rumour spreads over the entire lattice. We conjecture that  similar  results hold for $p > 0$ as well.

For the sake of completeness, Fig. \ref{Fig_R0} shows the influence of the basic reproduction number $\mathcal{R}_0 = \lambda / \alpha$ on $\langle R \rangle $. As expected, provided that $\mathcal{R}_0 > 1$  and that $Q$ is small enough such that the absorbing configurations of the cultural dynamics are monocultural, the rumour spreads to a sizeable portion of the lattice, but never to the entire lattice as pointed out before. The transition at $\mathcal{R}_0 = 1$ is continuous \cite{daley1964epidemics,maki1973mathematical}. However, for large $Q$  such that the absorbing configurations of the cultural dynamics are multicultural, the rumour reaches only a few agents close to the location of the initial spreader, regardless of the value of  $\mathcal{R}_0$.

\section{Conclusion}

In his seminal book, Rogers has offered  plentiful empirical evidence that cultural similarity plays a major role in the fate of innovations \cite{thirdedition}: innovations are more likely to spread among individuals sharing similar  beliefs.  The understanding of the role of heterogeneity in the spreading of innovations  has allowed companies to reduce  risk and maximize the impact of new products. 
In this paper, we add to Roger's conjecture by exploring a scenario in which the cultural heterogeneity of the community is described by Axelrod's model of culture dissemination and the diffusion of the innovation is described by a variant of the Daley and Kendall  model of rumour propagation.  Hence the interchangeability of the words innovation and rumour throughout the paper.

\textcolor{rev}{
Our results suggest that curbing the spread of rumours does not necessarily require changes in the social fabric, i.e., changes in the topology of the agents interaction networks \cite{holme2006nonequilibrium, fu2008coevolutionary, iniguez2009opinion}. 
In fact, this outcome appears naturally  in communities of heterogeneous agents who are allowed to hold different opinions and beliefs, thus providing theoretical support for Roger's empirical observations.
Specifically, }we find that cultural heterogeneity always impairs the propagation of the rumour and that, beyond a critical level of heterogeneity, the rumour is stuck in the neighborhood of its creator.

\textcolor{rev}{
Rogers' theory of the diffusion of innovations is a sociological theory that builds on  many natural experiments (e.g., adoption of hybrid corn,  Norplant 	contraceptive, rap,  cellular phones and  Nintendo home video-game players) to determine the characteristics of both the innovation and  the target community that best correlate with the success  of the  innovation \cite{thirdedition}. The theory contends  that the number of adopters of an innovation increases in time following an S-shaped growth curve (see \cite{Tilles2015} for a study of these growth curves for the pure Axelrod dynamics). Since our entire analysis is based on the characterization  of the absorbing configurations, we have access only to the ultimate number of adopters of the innovation, viz., $\langle R \rangle$.  In that sense, a direct comparison with real data is not possible, as those data focus solely on the growth of the number of adopters. Nevertheless, our results  support Rogers' conjectures and empirical observations in a qualitative way. For instance, Rogers' discussion of  homophily as a barrier to the diffusion of an innovation is in full agreement with the conclusions of our study.
}

\textcolor{rev}{The study of the coupled culture-rumour dynamics adds to the rapidly growing literature on social physics \cite{jusup2022social}. In fact, the availability of high quality data on social patterns and human activity (see, e.g., \cite{sociopatterns}), as well as the great appeal and  relevance of social issues,  has  prompted physicists' contributions   to several interdisciplinary areas, including (but not limited to) collective intelligence \cite{reia2019agent, perc2017statistical}, cultural polarization \cite{baumann2020modeling}, criminal behavior \cite{d2015statistical}, human mobility \cite{gonzalez2008understanding}, urban dynamics \cite{louf2013modeling, reia2022spatial, reia2022modeling}, traffic flow \cite{nagatani2002physics, schadschneider2002traffic}, social networks \cite{watts1998collective, goh2007human}, and moral behavior \cite{capraro2019,Fontanari2023}. These advances have been made possible by the maturing of complex systems as a research field, as attested by the 2021 Nobel Prize in Physics \cite{bianconi2023complex}. Perhaps the key characteristic of a complex system is the emergent phenomena resulting from its interacting components. In the coupled culture-rumour dynamics, the emergent phenomenon is the discontinuous phase transition that separates the regime where the rumour  spreads over the community  from the regime where it dies out close to its creator. The existence of  this transition could only be revealed through a mathematical and computational modeling, hence the importance of assessing Roger's theory for the diffusion of innovations using the tools of statistical mechanics.
}

\acknowledgments
JFF was partially supported by Funda\c{c}\~ao de Amparo \`a Pesquisa do Estado de S\~ao Paulo (FAPESP), grant 2020/03041-3 and  Conselho Nacional de Desenvolvimento Cient\'{\i}fico e Tecnol\'ogico (CNPq),  grant  305620/2021-5. AMP acknowledges the support of S\~ao Paulo Research Foundation (FAPESP), grant
2019/22277-0. 
This research was conducted using the computational resources of the  Center for Research in Mathematical Sciences Applied to Industry (CeMEAI) funded by FAPESP, grant 2013/07375-0.

\bibliography{references}
\bibliographystyle{eplbib}

\end{document}